\documentclass{llncs}

\usepackage[T1]{fontenc}
\usepackage{lmodern}
\usepackage[scaled=.8]{beramono}
\usepackage{latexsym}
\usepackage{textcomp}
\usepackage{paralist}
\usepackage[utf8]{inputenc}
\usepackage{hyperref}

\usepackage[english]{babel} 
\usepackage{amssymb}
\usepackage{cite}
\usepackage{listings}
\usepackage[linesnumbered,lined,boxed,commentsnumbered]{algorithm2e}
\usepackage{amsfonts}
\usepackage{pdflscape}
\usepackage{afterpage}
\usepackage{capt-of}% or use the larger `caption` package
\usepackage{lipsum}% dummy text

\usepackage{algpseudocode}
\algnewcommand\algorithmicswitch{\textbf{switch}}
\algnewcommand\algorithmiccase{\textbf{case}}

\usepackage{mathabx}
\newtheorem{requirement}{Requirement}

\usepackage{url}
\newcommand{\keywords}[1]{\par\addvspace\baselineskip
\noindent\keywordname\enspace\ignorespaces#1}

\usepackage{graphicx}
\usepackage{enumerate}
\usepackage{caption}
\usepackage{subcaption}
\usepackage{multirow}
\usepackage{array}
\usepackage{booktabs}
\usepackage{todonotes}

\usepackage{enumitem}
\usepackage{paralist}
 \title{Co-evolution of RDF Datasets}%: Challenges and Solutions
 \author{ Sidra Faisal, Kemele M. Endris, Saeedeh Shekarpour, S\"{o}ren Auer, Maria-Esther Vidal}
\institute{University of Bonn \& Fraunhofer IAIS, Bonn, Germany\\
\email{\{lastname\}@cs.uni-bonn.de} }

\begin{document}

\maketitle

\begin{abstract}
Linking Data initiatives have fostered the publication of large number of RDF datasets in the Linked Open Data (LOD) cloud,  as well as the development of  query processing infrastructures to access these data in a federated fashion. 
However, different experimental studies have shown that availability of  LOD datasets cannot be always ensured, being RDF data replication required for envisioning {\it reliable} federated query frameworks.   
%For many use cases it is not feasible to access RDF data in a truly federated fashion.
%For consistency, latency and performance reasons data needs to be replicated in order to be used locally.
%However, both a replica and its origin dataset undergo changes over time.
Albeit enhancing data availability, RDF data replication requires synchronization and conflict resolution when replicas and source datasets are allowed to change data over time, i.e., {\it co-evolution} management needs to be provided to ensure consistency. 
%The concept of co-evolution refers to mutual propagation of the changes between a replica and its origin dataset.
%The co-evolution process addresses synchronization and conflict resolution issues.
In this paper, we tackle the problem of RDF data co-evolution and devise an approach for conflict resolution during co-evolution of RDF datasets. 
% In this article, we initially provide formal definitions of all the concepts required for realizing co-evolution of RDF datasets.
%Then, we propose a methodology to address the co-evolution of RDF datasets. 
Our proposed approach is {\it property-oriented} and allows for exploiting {\it semantics} about RDF properties during  co-evolution management.
%We rely on a property-oriented approach for employing the most suitable strategy or functionality.
The quality of our approach is empirically evaluated in different scenarios on the DBpedia-live dataset. 
%This methodology was implemented and tested for a number of different scenarios.
%The result of our experimental study shows the performance and robustness aspect of this methodology.
Experimental results suggest that proposed proposed techniques have a {\it positive impact} on the quality of  data in source datasets and replicas. 
\end{abstract}

\keywords{Dataset Synchronization, Dataset Co-evolution, Conflict Identification, Conflict Resolution, RDF Dataset}

\section{Introduction}
During the last decade, the Linked Open Data (LOD) cloud has considerably grown~\cite{DBLP:conf/semweb/SchmachtenbergBP14}, comprising currently more than 85 billion triples from approximately 3400 datasets\footnote{Observed on 17th December 2015 on \url{http://stats.lod2.eu/}.}. 
Further,  Web based interfaces such as SPARQL endpoints~\cite{SPARQLProtocol} and Linked Data fragments~\cite{VerborghHMHVSCCMW14}, have been developed to access RDF data following the HTTP protocol, while federated query processing frameworks allow users to pose queries against federations of RDF datasets. 
Nevertheless, empirical studies by Buil-Aranda et al.~\cite{DBLP:conf/semweb/ArandaHUV13} suggest the lack of Web availability of a large number of LOD datasets, being frequently required the replication of  small portions of data, i.e.,  slices of an RDF dataset,  to enhance reliability  and performance of Linked Data applications~\cite{DBLP:conf/semweb/EndrisFOAS15}. 
Although RDF replication allows for enhancing RDF data availability,  synchronization problems may be generated because source datasets and replicas {\it may change} over time,  e.g., \emph{DBpedia Live mirror tool}\footnote{\url{https://github.com/dbpedia/dbpedia-live-mirror}} publishes changes in a public changesets folder\footnote{\url{http://live.dbpedia.org/change sets/}}. 
%Recently, the amount of structured data which has been published on the Web as Linked Open Data (LOD) has been enormously grown.
%Currently, it comprises more than 85 billion triples from approximately 3400 datasets\footnote{observed on 17th December 2015 on \url{http://stats.lod2.eu/}.}. 
%Still, mainly because of performance issues, small applications prefer to have a small replica (i.e., a slice of RDF data) for their local data consumption.
%Both a replica and its origin dataset undergo changes over time.
%Huge RDF datasets basically publish the updates as different RDF files that contain only the changes (i.e. removed and added triples);
%hese can also be downloaded and easily integrated to the older versions of the dataset.  
%For instance, \emph{DBpedia Live mirror tool}\footnote{\url{https://github.com/dbpedia/dbpedia-live-mirror}} publishes changes in a public changesets folder\footnote{\url{http://live.dbpedia.org/change sets/}}. 

{\it Co-evolution} refers to mutual propagation of the changes between a replica and its origin or source dataset, where propagation specially in a mutual way,  raises synchronization issues which need to be addressed to avoid data inconsistency.
Issues are about how changes should be propagated and in case of {\it inconsistencies} or {\it data conflicts}, how these conflicts should be resolved.
Thus, our main research problem is to develop a co-evolution process able to exploit the properties of RDF data and solve conflicts generated by the propagation of changes among source datasets and replicas.  We propose  a two-fold co-evolution approach, comprised of the following components:
\begin{inparaenum}[\itshape i\upshape)]
\item an RDF data  synchronization component, and 
\item a component for conflict identification and resolution.
\end{inparaenum}

Our approach relies on the {\it assumption} that either the source dataset provides a tool to compute a changeset at real-time or  third party tools can be used for this purpose. 
%In this article, the origin dataset is called source dataset and the replica is called target dataset.
%We assume that either the source dataset integrates a tool to compute a changeset at real-time or the third party tools can be used for this purpose. 
Another {\it assumption} is that {\it slices} of the RDF data from the source dataset are replicated in the replicas or {\it target datasets}, where a slice\footnote{An RDF slice is also known as a fragment in the approaches proposed by Iba\~nez et al.~\cite{Ibanez2014}, Montoya et al.~\cite{MontoyaISWC2015}, and  Verborgh et al.~\cite{VerborghHMHVSCCMW14}.} corresponds to an RDF subgraph of the source RDF graph~\cite{SALEEM2013}.

\begin{figure}[tb]
\centering
\includegraphics[width=0.7\textwidth]{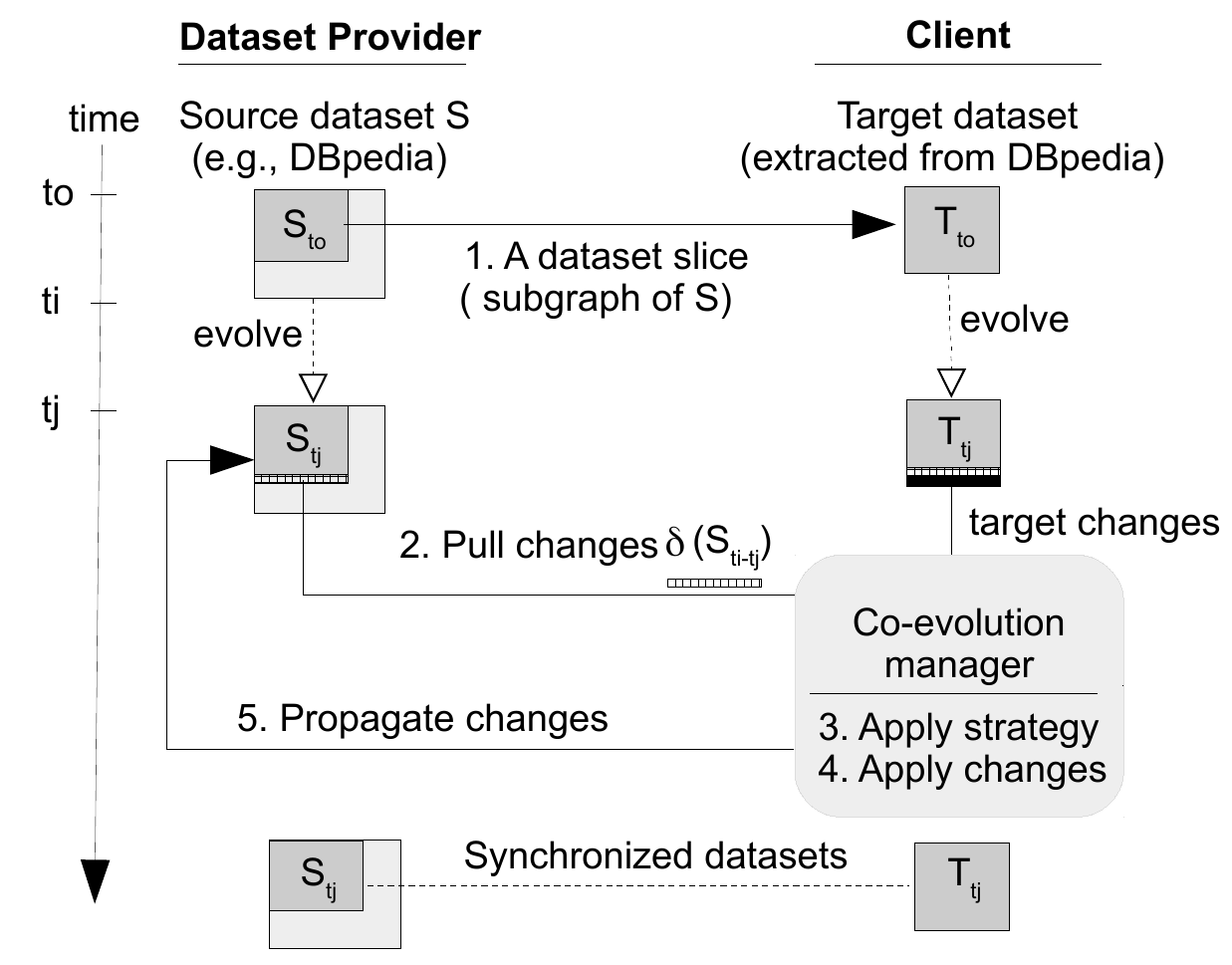}
\caption{Co-evolution of linked datasets}
\label{fig:co-evolution}
\vspace{-1.75em}
\end{figure}

Figure\ref{fig:co-evolution} illustrates the co-evolution between two RDF datasets. 
Initially, a slice of source dataset is used to create a target dataset, i.e.,   the target dataset $T_{t_0}$ is sliced from the source dataset $S_{t_0}$ of dataset $S$ at time $t_0$.
Both the source and target datasets evolve themselves with the passage of time, e.g., these datasets evolve to $S_{t_j}$ and $T_{t_j}$ during timeframe $ t_i-t_j $, while $ t_i \, < \, t_j $. 
Changes from $ S_{t_j} $, denoted by $ \delta(S_{t_i-t_j})$,  are propagated to the target   and vice versa by the RDF data  synchronization component.
For synchronization, changes from both source and target datasets are compared to identify conflicts.
The resolved conflicts are applied on the source and target datasets to vanish inconsistencies, for example, at time point $t_j$, the co-evolution manager identifies the conflicts and resolves them.
The conflicts are resolved and final changes are merged in both datasets.  
 
We empirically evaluate the quality of our co-evolution approach on different co-evolution scenarios of data from the  DBpedia\footnote{\url{http://wiki.dbpedia.org/}} and changesets from DBpedia-live published from September 01, 2015 to October 31, 2015 using iRap~\cite{irap2015}.
The goal of the evaluation is to study the impact on data quality of the propose co-evolution process, where quality is measured in terms of completeness, consistency, and consciseness~\cite{Zaveri2015}.
Observed experimental results suggest that our synchronization, and conflict identification and resolution techniques positively affect the quality of the data in both the source and target datasets. 
 
The paper is structured as follows: 
~\autoref{sec:preliminaries} provides formal definitions of the basic notations and concepts used in the proposed co-evolution approach.
~\autoref{sec:problemstatement} presents detailed problem description and different synchronization strategies.
We then present the proposed approach in~\autoref{sec:approach} followed by evaluation in~\autoref{sec:evaluation}.
~\autoref{sec:relatedwork} presents the related work.
We close with the conclusion and the directions for the future work.

\section{Motivating example}		
\begin{figure}[tb]
\centering
\includegraphics[width=1.00\textwidth]{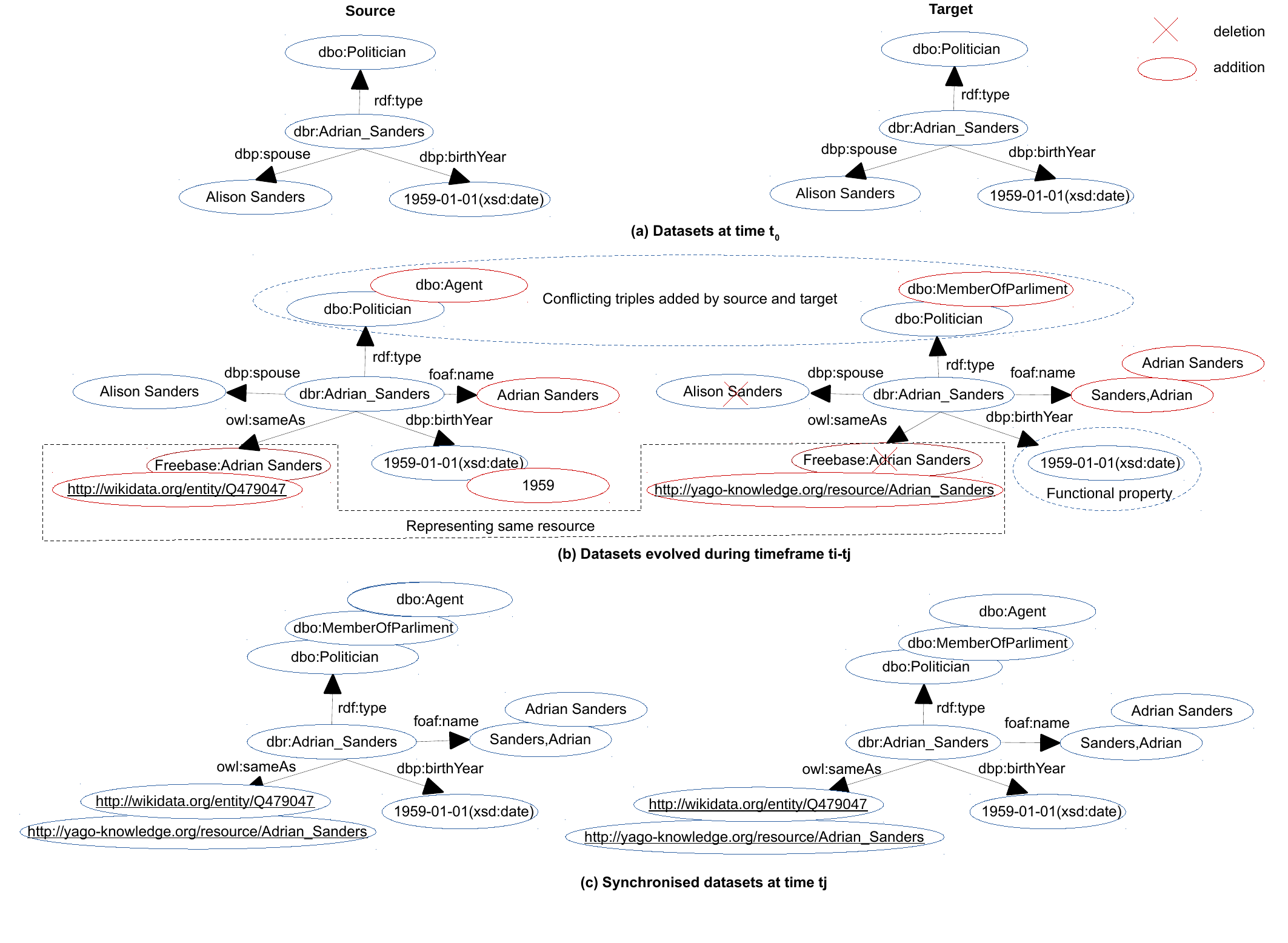}
\caption{Motivating example: a) Target dataset initialization, b) evolution, and c) synchronization with source}
\label{fig:motivation}
\end{figure}

%\paragraph{Use case scenario:} Herein, we present a use case scenario and track this scenario during the entire paper. 
Let us assume an application which requires information of politicians (e.g., name, birthYear, and spouse).
This information can be sliced from the datasets like DBpedia \footnote{http://dbpedia.org}, and used locally by the application. 
We use the following SPARQL query to slice DBpedia for our use case scenario:
\begin{lstlisting}
 CONSTRUCT  WHERE {
     ?s    rdf:type       dbo:Politician.
     OPTIONAL {
     ?s    foaf:name      ?name.
     ?s    dbp:birthYear  ?birthYear.
     ?s    dbp:spouse     ?spouse.
     ?s    owl:sameAs     ?sameAs  }
     }
 \end{lstlisting}
 
Our approach is inspired from the scenario described in~\autoref{fig:motivation}. 
Initially, at time $t_0$, this slice is used to populate target dataset.
Both source and target datasets evolve during timeframe $ t_i-t_j $, while $ t_i \, < \, t_j $. 
Source dataset adds object value $dbo:Agent$ for rdf:type, $Adrian Sanders$ for foaf:name, $1959$ for dbp:birthYear, and $Freebase:Adrian Sanders$ and \url{http://wikidata.org/entity/Q479047} for owl:sameAs to resource dbr:Adrian\_Sanders. 
Target dataset adds object value $dbo:MemberOfParliment$ for rdf:type, $Sanders,Adrian$ for foaf:conname, and $Freebase:Adrian Sanders$ and \url{http://yago-knowledge.org/resource/Adrian\_Sanders} for owl:sameAs to resource dbr:Adrian\_Sanders.  

For resource dbr:Adrian\_Sanders, we have two different values for rdf:type in source and target changesets.
We need to check which of them is correct.
We already know dbr:Adrian\_Sanders can be an agent and member of parliment at the same time. 
However, this check can be made by looking whether the two classes are disjoint or not. Source adds object value $1959$ for dbp:birthYear to dbr:Adrian\_Sanders. 
As dbp:birthYear is a functional property, it can have only one value.
So, we have to choose one value among the already existing value $1959-01-01$ in dataset and the new value $1959$ in the changeset.
One solution can be to randomly select one value among two.
Similarly, source adds object value $Freebase:Adrian Sanders$ for owl:sameAs while target dataset deletes this value after adding it. 
Considering target as a more customized dataset, we prefer the changes of target over source changes. 
Thus, we delete $Freebase:Adrian Sanders$ in synchronized dataset. 
We still have two different owl:sameAs values for dbr:Adrian\_Sanders. 
However, as they are representing the same resource, we will keep both values in synchronized dataset.   

\section{Preliminaries}\label{sec:preliminaries}\label{sec:problemstatement}
In this section, we formalize the main concepts required for realizing co-evolution of RDF datasets. 
The \emph{Resource Description Framework (RDF)}\footnote{\url{http://www.w3.org/TR/rdf11-concepts/}} is widely used to represent information on the Web. 
A resource can be any thing (either physical or conceptual). 
The RDF data model expresses statements about Web resources in the form of subject-predicate-object (triple).
The subject denotes a resource; the predicate expresses a property of subject or a relationship between the subject and the object; the object is either a resource or literal. 
For identifying resources, RDF uses Uniform Resource Identifiers (URIs)\footnote{A URI is a string of characters used as unique identifier for a Web resource.}
and Internationalized Resource Identifier (IRIs)\footnote{A generalization of URIs enabling the use of international character sets.}.
The rationale behind is that the names of resources must be universally unique.
We assume that both source and target datasets are RDF datasets. An RDF dataset is formally defined as follows:

\begin{definition}[RDF Dataset]
Formally, an RDF dataset is a finite set of triples $(s, p, o) \in (I \cup B) \times I \times (I \cup L \cup B)$, where $I, B, and L$ are the disjoint sets of all IRIs, blank nodes, and literals \cite{irap2015}.
%Formally, an RDF dataset is a finite set of triples $(s, p, o) \in (R  \cup B) \times P \times (R \cup L \cup B)$, where $R$ is the set of all RDF resources, $B$ is the set of all blank nodes ($B \cap R = \emptyset$), $P$ is the set of all predicates ($P \subseteq R$) and $L$ is the set of all literals ($L \cap R = \emptyset$).
\end{definition}

%\paragraph{Use case scenario:} Herein, we present an use case scenario and track this scenario during the entire paper. 
%Let us assume a mobile application which requires information of restaurants (i.e., name, rating, chef, abstract, and depiction) nearby users' location.
%%At first, the required data can be extracted from LinkedGeoData or DBpedia that is a slice we dubbed as 'Target dataset'. 
%This information can be sliced from the huge datasets like LinkedGeoData\footnote{http://linkedgeodata.org} or DBpedia \footnote{http://dbpedia.org} and use locally by the mobile application. 
%We use the following SPARQL query to slice DBpedia for our use case scenario:
%\begin{lstlisting}
% CONSTRUCT 
% WHERE {
%     ?s    a               dbo:Restaurant.
%     ?s    rdfs:label      ?label.
%     ?s    georss:point    ?point.
%     ?s    dbo:abstract    ?abstract.
%     ?s    dbp:rating      ?rating.
%     ?s    foaf:depiction  ?depiction.
%     ?s    dbp:headOfChief ?headOfChief
%  }
% \end{lstlisting}
Let us assume that the slice contains the following triples%~\footnote{@prefix wmcf: http://commons.wikimedia.org/wiki/Special:FilePath/}
 %running the SPARQL query yields the following triples:
\begin{lstlisting}[caption={Content of initial target dataset}, captionpos=b, label={lst:targetinit}]
dbr:Adrian_Sanders rdf:type      dbo:Politician;
		   dbp:spouse    Alison Sanders;
		   dbp:birthYear 1959-01-01 (xsd:date).
\end{lstlisting}
%dbr:FLIP_Burger_Boutique   a     dbo:Restaurant;
%         foaf:name         "FLIP"@en ;
%         georss:point      "33.7984 -84.4159";
%         dbo:abstract      "FLIP is an upscale full-service American restaurant ... "@en;
%         dbp:rating        4;
%         foaf:depiction    wmcf:Richard_Blais.JPG ;
%         dbp:headChef      "Richard Blais"@en.
%dbr:Jean_Georges  a        dbo:Restaurant;
%         foaf:name         "Jean-Georges Vongerichten"@en;
%         georss:point      "40.76905277777778 -73.98143055555556";
%         dbo:abstract      "Jean Georges is a three-Michelin-stars restaurant at 1 Central Park ..."@en;
%         dbp:rating        5;
%         foaf:depiction    wmcf:JGV.jpg;
%         dbp:headChef      "Mark LaPico"@en.

This local copy of sliced dataset, referred as {\it target dataset}, might undergo changes by user feedback (e.g. user can update the restaurant rating or fulfil abstract information).
% Then, the user of the app can add new restaurant information or update the existing one, causing the initial target dataset to evolve. 
After some time, DBpedia dataset also evolves by adding new restaurants information or updating the existing ones. 
As a result, {\it target dataset} might be out of date and need to be synchronized with DBpedia. 
During synchronization, a conflict (defined in \autoref{def:conflict}) might occur, if the same information was updated by the source (DBpedia) dataset and the target dataset (by the app users). 

\begin{definition}[Evolving RDF Dataset]\label{def:evolvingDataset}
Let us assume that $D_{t_i}$ represents the version of the RDF dataset $D$ at the particular time $t_i$.
An evolving dataset $D$ is a dataset whose triples change over time. 
In other words, for timeframe $t_i-t_j$,  there is a triple $x$ such as either $( x \in D_{t_i}  \wedge x \notin D_{t_j}) $ or $ (x \notin D_{t_i}  \wedge x \in D_{t_j} )$.
\end{definition}

\begin{definition}[Changeset]\label{def:changeset}
Let us assume that $D$ is an evolving RDF dataset. 
and $D_{t_i}$ is the version of $D$ at time $t_i$. 
A changeset which is denoted by $\delta ( D_{t_i-t_j} ) $ shows the difference of two versions of an evolving RDF dataset in a particular timeframe $ t_i-t_j $, while $ t_i \, < \, t_j $.
The changeset is formally defined as $ \delta ( D_{t_i-t_j} ) = < \delta ( D_{t_i-t_j} ) ^ + , \delta ( D_{t_i-t_j} ) ^ - > $ where,
\begin{compactitem}
\item $ \delta ( D_{t_i-t_j} ) ^ + $ is a set of triples which have been added to the version $D_{t_j}$ in comparison to the version $D_{t_i}$. 
\item $ \delta ( D_{t_i-t_j} ) ^ - $ is a set of triples which have been deleted from the version $D_{t_j}$ in comparison to the version $D_{t_i}$. 
\end{compactitem}
\end{definition}

\begin{example}[Changesets]\label{ex:changeset}
Let the following files are found as changesets at time $t_i$ from the source and target datasets.

%dbr:FLIP_Burger_Boutique  foaf:name       "FLIP"@en ;
%                          dbo:abstract    "FLIP is an upscale full-service .."@en.
%dbr:Jean_Georges          foaf:name       "Jean-Georges Vongerichten"@en;
%                          foaf:depiction  wmcf:JGV.jpg.
\begin{lstlisting}[caption={Source changeset, (A)=$ \delta ( S_{t_i-t_j} ) ^ -$ , and (B) = $ \delta ( S_{t_i-t_j} ) ^ +$ }, captionpos=b,label={lst:sourcechangesets} ]
#(A). Deleted triples
#______________________________________________________________________________________
#(B). Added triples
dbr:Adrian_Sanders rdf:type      dbo:Agent;
		   foaf:name     Adrian Sanders;
		   dbp:birthYear 1959;
		   owl:sameAs    Freebase:Adrian Sanders;
		   owl:sameAs    http://wikidata.org/entity/Q479047.
\end{lstlisting}
%dbr:FLIP_Burger_Boutique  foaf:name       "FLIP burger boutique"@en ;
%                          dbo:abstract 	  "FLIP burger boutique (stylized as FLIP).."@en.
%dbr:Jean_Georges          foaf:depiction  wmcf:FLIP.jpg;
%                          foaf:name       "Jean-Georges"@en.
%dbr:Het_Groot_Paradijs    a               dbo:Restaurant;
%                          foaf:name       "Het Groot Paradijs";
%                          dbo:abstract    "Het Groot Paradijs was a restaurant located in Middelburg,..";
%                          georss:point    "51.50027222222222 3.6166805555555555";
%                          dbp:rating      4.

%dbr:Jean_Georges          foaf:name      "Jean-Georges Vongerichten"@en;
%                          foaf:depiction  wmcf:JGV.jpg.

\begin{lstlisting}[caption={Target changeset, (A)= $ \delta ( T_{t_i-t_j} ) ^ -$ , and (B) = $ \delta ( T_{t_i-t_j} ) ^ +$}, captionpos=b,label={lst:targetchangesets}]
#(A) Deleted triples
dbr:Adrian_Sanders dbp:spouse    Alison Sanders;
		   owl:sameAs    Freebase:Adrian Sanders.                          
#______________________________________________________________________________________
#(B) Added triples
dbr:Adrian_Sanders rdf:type      dbo:MemberOfParliment;
		   foaf:name     Adrian Sanders;
		   foaf:name     Sanders, Adrian;
		   owl:sameAs    Freebase:Adrian Sanders;
		   owl:sameAs    http://yago-knowledge.org/resource/Adrian_Sanders.
\end{lstlisting}
\end{example}						  

%dbr:Jean_Georges          foaf:name      "JGV restaurant"@en;
%                          foaf:depiction  myp:jgvn.jpg.

\begin{definition}[Synchronized Dataset]\label{def:syncds}
Two evolving datasets, $D^{(1)}$ and $D^{(2)}$, are said to be synchronized (or in sync) iff one of the following is true at a given time $t_k$: i) $D^{(1)}_{t_k} \subseteq D^{(2)}_{t_k}$, ii) $D^{(2)}_{t_k} \subseteq D^{(1)}_{t_k}$, or iii) $D^{(1)}_{t_k} \equiv D^{(2)}_{t_k}$.
\end{definition}

\section{Problem Statement}\label{sec:problemstatement}
The core of the co-evolution concept relies on the mutual propagation of changes between the source and target datasets in order to keep the datasets {\it in sync}.
Thus, from time to time, the target dataset and the source dataset have to exchange the changesets and then update the local repositories.
Updating a dataset with changesets from the source dataset might cause inconsistencies.
%Please note that updating a dataset with changesets from the source dataset might cause inconsistencies. 
Our co-evolution strategy aims at dealing with changesets from either the source or target dataset and provide a suitable reconciliation strategy. 
Various strategies can be employed for synchronising datasets.
In this section we provide requirements and formal definitions for guiding the co-evolution process.

\subsection{Synchronization}
In the beginning the target dataset is derived (as a slice or excerpt) from the source dataset, thus the following requirement always holds.

\begin{requirement}[Initial Inclusion]\label{req:Initial_Inclusion}
At the initial time $t_0$, the target dataset $T$ is a subset of the source dataset S: $ T_{t_0} \subseteq S_{t_0} $, and thus source and target datasets are in sync.
\end{requirement}

After some time, both source and target datasets evolve. 
At time $t_i$, the target dataset is $T_{t_i} = T_{t_0} \cup \delta (T_{t_0-t_i})$ and the source dataset is $S_{t_i} = S_{t_0} \cup \delta (S_{t_0-t_i})$.
%Synchronization strategies should be used when $ \delta (T_{t_0}) \neq \delta ( S_{t_0} )$. 

\begin{requirement}[Required Synchronization]\label{req:Required_Synchronization}
At time $t_j$, a synchronization of both datasets is required iff source and target datasets were synchronised at time $t_i$, and the changesets applied to source and target datasets differ, i.e. $ \delta ( S_{t_i-t_j}) \neq \delta (T_{t_i-t_j})$.
\end{requirement}

\subsection{Conflict}
When we synchronize the target $T_{t_i}$ with source $S_{t_i}$, there may exist triples which have been changed in both datasets. 
These changed triples may be conflicting.

\begin{definition}[Potential Conflict]\label{def:conflict}
Let us assume that a synchronization is required for a given time slot $t_i-t_j$. $ \delta (S_{t_i-t_j}) $ is the changeset of the source dataset and  $\delta (T_{t_i-t_j})$ is the changeset of the target dataset. 
A potential conflict is observed when there are triples $x_1 = ( s, \, p, \, o_1) \in S_{t_j} \wedge x_2 = ( s, \, p, \, o_2)  \in \delta (T_{t_i-t_j}) \wedge x_2 \notin S_{t_j} =S_{t_i} \cup \delta (S_{t_i-t_j}) $ with $o_1 \notequiv o_2$.
\end{definition}

Taking $o_1 \notequiv o_2 $ as an indication for a conflict is subjective; in the sense that the characteristics of the involved property $p$ influences the decision.
Consider two triples $ (s, \, p, \, o_1) $ and $ (s, \, p, \, o_2) $. 
If $p$ is a functional data type property, two triples are conflicting iff the object values $o_1$ and $o_2$ are not equal.
However, if the property $p$ is a functional object property, these two triples are conflicting if the objects are or can be inferred to be different (e.g. via \verb|owl:differentFrom|).
%will not be conflicting in case there exists some triples such as $ (o_1 \; owl:sameAs \; o_2) $ or $ (o_2 \; owl:sameAs \; o_1) $. 
%Otherwise, they will be conflicting as a functional property have exactly one object value against a subject.  
%If $p_1$ is owl:sameAs, these triples will not be treated as conflicting.  
Another property which needs special consideration is \verb|rdf:type|.
For this property it is necessary to check whether $o_1$ and $o_2$ belong to disjoint classes. 
Only then these triples would be conflicting. 
For example, \verb|s1 rdf:type Person|  and \verb|s1 rdf:type Athlete| are not conflicting if \verb|Athlete| is a subclass of \verb|Person| (i.e. not disjoint).  
Thus, the process of detecting conflicts is considering the inherent characteristics of the involved property.

\subsection{Synchronization Strategies}\label{def:strategies}
In the following, we list possible strategies for synchronization.
We consider the time frame $t_i-t_j$, where in the time $t_i$, the source and target datasets are synchronised and until time $t_j$, both source and target datasets have been evolving independently.
Before applying synchronization, the state of the source dataset is $S_{t_j}=S_{t_i} \cup \delta (S_{t_i-t_j})$ and the target dataset is $T_{t_j}=T_{t_i} \cup \delta (T_{t_i-t_j})$.

\subsubsection{Strategy I:}
This synchronization strategy prefers the source dataset and ignores all local changes on the target dataset; thus, the following requirement is necessary. 

\begin{requirement}[Inclusion for synchronization]\label{req:Inclusion1}
At any given time $t_j$, after synchronising using selected strategy, the target dataset should be a subset of the source dataset, i.e. $T_{t_j} \subseteq S_{t_j} $.
\end{requirement}

Therefore, the target dataset ignores all triples $ \{ x \, | \, x \notin \delta (S_{t_i-t_j}) \, \wedge \, x \in \delta (T_{t_i-t_j}) \} $ and adds only the triples $ \{ y \, | \, y \in \delta (S_{t_i-t_j}) \} $.
After synchronization, the state of source dataset is $S_{t_j}=S_{t_i} \cup \delta (S_{t_i-t_j})$ and the state of the target dataset is $T_{t_j}=T_{t_i} \cup \delta (S_{t_i-t_j})$.
Thus, the requirement \autoref{req:Inclusion1} is met and $ T_{t_j} \subseteq S_{t_j} $. 
A special case of this strategy is when the target is not evolving. 

\begin{example} \label{exmp:strategyI}
Applying strategy I for synchronization on \autoref{ex:changeset} gives the following triples:
%frame=none,xleftmargin=\parindent, basicstyle=\footnotesize\ttfamily, 

\begin{lstlisting}[breaklines=true]
dbr:Adrian_Sanders rdf:type      dbo:Politician;
		   rdf:type      dbo:Agent;
		   foaf:name     Adrian Sanders;							  
		   dbp:spouse    Alison Sanders;
		   dbp:birthYear 1959-01-01 (xsd:date);
		   dbp:birthYear 1959;
		   owl:sameAs    Freebase:Adrian Sanders;
		   owl:sameAs    http://wikidata.org/entity/Q479047.						  
\end{lstlisting}
\end{example}

%dbr:FLIP_Burger_Boutique   a     dbo:Restaurant;
%	 foaf:name          "FLIP burger boutique"@en ;
%	 dbo:abstract 	    "FLIP burger boutique (stylized as FLIP).."@en.
%         georss:point       "33.7984 -84.4159";
%         dbp:rating         4;
%         foaf:depiction     wmcf:Richard_Blais.JPG ;
%         dbp:headChef       "Richard Blais"@en.
%dbr:Jean_Georges  a         dbo:Restaurant;
%	 foaf:depiction     wmcf:FLIP.jpg;
%         foaf:name          "Jean-Georges"@en.
%         georss:point       "40.76905277777778 -73.98143055555556";
%         dbo:abstract       "Jean Georges is a three-Michelin-stars restaurant at 1 Central Park ..."@en;
%         dbp:rating         5;
%         dbp:headChef       "Mark LaPico"@en.
%dbr:Het_Groot_Paradijs      a       dbo:Restaurant;
%         foaf:name          "Het Groot Paradijs";
%         dbo:abstract       "Het Groot Paradijs was a restaurant located in Middelburg,..";
%         georss:point       "51.50027222222222 3.6166805555555555";
%         dbp:rating         4.
         
\subsubsection{Strategy II:}
With this strategy, the target dataset is not synchronized with the source dataset and keeps all its local changes.
Thus, the target dataset is not influenced by any change from the source dataset and evolves locally.
After synchronization, at time $t_j$, the state of the target dataset is $T_{t_j}=T_{t_i} \cup \delta (T_{t_i-t_j}) $, and the state of the source dataset is $S_{t_j}=S_{t_i} \cup \delta (S_{t_i-t_j}) $.	
It allows for synchronized replicas only if data is deleted.
There is no synchronization if triples in the target dataset are updated or new triples are included.
\begin{example} \label{exmp:strategyII}
Applying strategy II for synchronization on \autoref{ex:changeset} gives the following triples:

\begin{lstlisting}[breaklines=true]
dbr:Adrian_Sanders rdf:type      dbo:Politician;
		   rdf:type      dbo:MemberOfParliment;
		   foaf:name     Adrian Sanders;
		   foaf:name     Sanders, Adrian;							  
		   dbp:birthYear 1959-01-01 (xsd:date);                    
		   owl:sameAs    http://yago-knowledge.org/resource/Adrian_Sanders.
\end{lstlisting}
\end{example}

%dbr:FLIP_Burger_Boutique   a     dbo:Restaurant;
%         foaf:name         "FLIP"@en ;
%         georss:point      "33.7984 -84.4159";
%         dbo:abstract      "FLIP is an upscale full-service American restaurant ... "@en;
%         dbp:rating        4;
%         foaf:depiction    wmcf:Richard_Blais.JPG ;
%         dbp:headChef      "Richard Blais"@en.
%dbr:Jean_Georges  a        dbo:Restaurant;
%         foaf:name         "JGV restaurant"@en;
%         georss:point      "40.76905277777778 -73.98143055555556";
%         dbo:abstract      "Jean Georges is a three-Michelin-stars restaurant at 1 Central Park ..."@en;
%         dbp:rating        5;
%         foaf:depiction    myp:jgvn.jpg;
%         dbp:headChef      "Mark LaPico"@en.

\subsubsection{Strategy III:} \label{strategy III} 
This synchronization strategy respects the changesets of both source and target datasets except that it ignores conflicting triples.
%\begin{requirement}[Strategy III - Inclusion]\label{req:InclusionS2}
%At any given time $t_j$, after synchronising using Strategy III, the target dataset is a subset of the source dataset, $ T_{t_j} \subseteq S_{t_j} $.
%\end{requirement}

Here, the set of triples in which conflicts occur is $X = \{ x_1 =  ( s, \, p, \, o_1) \in S_{t_j} \wedge x_2 = ( s, \, p, \, o_2)  \in \delta (T_{t_i-t_j}) \wedge x_2 \notin S_{t_j} $ with $o_1 \notequiv o_2\} $\footnote{Set of conflicting triples selected after considering the inherent characteristics of the involved property. In rest of the paper, we say potential conflict a conflict, unless otherwise specified.}. 
With Strategy III, the set of conflicting triples $X$ is removed from the target dataset while the source changeset $\delta ( S_{t_i-t_j} )$ and the target changeset $\delta ( T_{t_i-t_j} )$ are added. 
After synchronization, the state of the source dataset is $S_{t_j}=(S_{t_i}  \cup  \delta (S_{t_i-t_j}) \cup \delta (T_{t_i-t_j}) )\setminus X$ and the state of the target dataset is $T_{t_j}=(T_{t_i} \cup \delta (T_{t_i-t_j}) \cup \delta (S_{t_i-t_j})) \setminus X$.
Thus, requirement \autoref{req:Inclusion1} is met. 

\begin{example} \label{exmp:main}
Applying strategy III for synchronization on \autoref{ex:changeset} gives the following triples:

\begin{lstlisting}[breaklines=true]
dbr:Adrian_Sanders rdf:type      dbo:Politician;
		   rdf:type      dbo:Agent;
		   rdf:type      dbo:MemberOfParliment;					  
		   owl:sameAs    http://wikidata.org/entity/Q479047;
		   owl:sameAs    http://yago-knowledge.org/resource/Adrian_Sanders.							 
\end{lstlisting}
\end{example}
%		   dbp:birthYear 1959-01-01 (xsd:date);		
		   
%
%dbr:FLIP_Burger_Boutique   a     dbo:Restaurant;
%         foaf:name         "FLIP burger boutique"@en;
%         georss:point      "33.7984 -84.4159";
%         dbo:abstract      "FLIP burger boutique (stylized as FLIP).."@en;
%         dbp:rating        4;
%         foaf:depiction    wmcf:Richard_Blais.JPG ;
%         dbp:headChef      "Richard Blais"@en.
%dbr:Jean_Georges  a        dbo:Restaurant;
%         georss:point      "40.76905277777778 -73.98143055555556";
%         dbo:abstract      "Jean Georges is a three-Michelin-stars restaurant at 1 Central Park ..."@en;
%         dbp:rating        5;
%         dbp:headChef      "Mark LaPico"@en.
%dbr:Het_Groot_Paradijs     a       dbo:Restaurant;
%         foaf:name         "Het Groot Paradijs";
%         dbo:abstract      "Het Groot Paradijs was a restaurant located in Middelburg,..";
%         georss:point      "51.50027222222222 3.6166805555555555";
%         dbp:rating         4.

\subsubsection{Strategy IV:} \label{strategy IV}
This synchronization strategy also respects the changesets of both source and target datasets. 
In addition, it includes conflicting triples after resolving the conflicts.

%\begin{requirement}[Strategy IV - Inclusion]\label{req:InclusionS4}
%At any given time $t_j$, after synchronising using the Strategy IV, the target dataset is a subset of the source dataset, $ T_{t_j} \subseteq S_{t_j} $.
%\end{requirement}
Here, we consider the set of triples in which conflict occurs as $X = \{ x_1 =  ( s, \, p, \, o_1) \in S_{t_j} \wedge x_2 = ( s, \, p, \, o_2)  \in \delta (T_{t_i-t_j}) \wedge x_2 \notin S_{t_j} $ with $o_1 \notequiv o_2\}$. 
%$X = \{ x_1= ( s_1, \, p_1, \, o_1 ), x_2= ( s_2, \, p_2, \, o_2 )| x_1 \in \delta ( T_{t_i-t_j} ) \wedge x_2 \in \delta ( S_{t_i-t_j} )  \wedge s_1 = s_2  \wedge \, p_1 = p_2  \wedge \, o_1 \notequiv o_2 \} $. 
The conflicts over these triples should be resolved. 
It can be resolved using some resolution policy as described in \cite{Bleiholder2006}.
\autoref{tab:resolutionfunctions} shows a list of various policies for resolving the conflicts.    
Conflict resolution results in a new set of triples called $Y$ whose triples are originated from $X$ but their conflicts have been resolved.
Then, this new set (i.e. $Y$) is added to the both source and target datasets.
After synchronization, the state of the source dataset is $S_{t_j}=((S_{t_i} \cup \delta (S_{t_i-t_j}) \cup  \delta (T_{t_i-t_j})) \setminus X) \cup Y$ and the state of target dataset is $T_{t_j}=((T_{t_i} \cup  \delta (T_{t_i-t_j}) \cup \delta (S_{t_i-t_j})) \setminus X) \cup Y$. 
Thus, requirement \autoref{req:Inclusion1} is met.           

\begin{table}[ht]
\caption{Conflict resolution policies and functions}
\scriptsize
 \begin{tabular}{|l|llll|}
    \hline \textbf{Category} & \textbf{Policy} & \textbf{Function} & \textbf{Type} & \textbf{Description} \\
      \hline
       \multirow{8}{*}{\parbox{0.1\linewidth}{Deciding}} 
       		& \parbox{0.13\linewidth}{Roll the dice} & Any & A & Pick random value.\\
            \cline{2-5}        			
            
       		& \parbox{0.12\linewidth}{Reputation} & \parbox{0.15\linewidth}{Best source} & A & \parbox{0.5\linewidth}{Select the value from the preffered dataset.} \\                    
             \cline{2-5}  
       
         & \parbox{0.12\linewidth}{Cry with the wolves} & \parbox{0.15\linewidth}{Global vote} & A & \parbox{0.5\linewidth}{Select the frequently occurring value for the respective attribute among all entities.}  \\                         		
       		 \cline{2-5}   
       		  
       		& \multirow{2}{*}{\parbox{0.12\linewidth}{Keep up-to-date}}
        				& First* & A & \parbox{0.5\linewidth}{Select the first value in order.} \\  
        				 \cline{3-5}
		    			&  & Latest* & A & \parbox{0.5\linewidth}{Select the most recent value.} \\
              \cline{2-5}        
                             
  	 		& \multirow{3}{*}{\parbox{0.12\linewidth}{Filter}} 
  		 		& Threshold* & A &  \parbox{0.5\linewidth}{Select the value with a quality score higher than a given threshold.} \\ 
        				 \cline{3-5}
				& & Best* & A & \parbox{0.5\linewidth}{Select the value with highest quality score.} \\  
        				 \cline{3-5}
       			& & TopN* & A & \parbox{0.5\linewidth}{Select the N best values.} \\     
                           
	 \hline
      \multirow{3}{*}{\parbox{0.1\linewidth}{Mediating}}
      		& \multirow{3}{*}{\parbox{0.12\linewidth}{Meet in the middle}}   		
 			& \parbox{0.2\linewidth}{Standard deviation, variance} & N & \parbox{0.5\linewidth}{Apply the corresponding function to get value.} \\ 
             \cline{3-5}  	                         
 			
 			& & \parbox{0.2\linewidth}{Average, median} & N & \parbox{0.5\linewidth}{Apply the corresponding function to get value.}  \\ 
             \cline{3-5} 

             & & Sum & N & \parbox{0.5\linewidth}{Select the sum of all values  as the resultant.}  \\                              
                          
    \hline 
  {\parbox{0.1\linewidth}{Conflict ignorance}}
       		& \parbox{0.12\linewidth}{Pass it on} & Concatenation & A & \parbox{0.5\linewidth}{Concatenate all the values to get the resultant.} \\  
   
         \hline
       \multirow{7}{*}{\parbox{0.1\linewidth}{Conflict avoidance}} 
           
       		& \multirow{4}{*}{\parbox{0.12\linewidth}{Take the information}}
    				& Longest & {\parbox{0.08\linewidth}{S, C, T}} & \parbox{0.5\linewidth}{Select the longest (non-NULL) value.}  \\           			            		 \cline{3-5} 
             	     & & Shortest & {\parbox{0.08\linewidth}{S, C, T}} & \parbox{0.5\linewidth}{Select the shortest (non-NULL) value.}  \\ 	 						\cline{3-5}
		    		& & Max & N & \parbox{0.5\linewidth}{Select the maximum value from all.}  \\ 
             		\cline{3-5}
		    		& & Min & N & \parbox{0.5\linewidth}{Select the minimum value from all.}\\                              
             \cline{2-5}  	             
      		& \multirow{3}{*}{\parbox{0.12\linewidth}{Trust your friends}} 
		& {\parbox{0.15\linewidth}{Choose depending*}} & A & \parbox{0.5\linewidth}{Select the value that belongs to a triple having a specific given value for another given attribute.} \\
             \cline{3-5}                    	 
		& & \parbox{0.15\linewidth}{Choose corresponding} & A & \parbox{0.5\linewidth}{Select the value that belongs to a triple whose value is already chosen for another given attribute.} \\    
		   \cline{3-5}             
         & & \parbox{0.15\linewidth}{Most complete*} & A & \parbox{0.5\linewidth}{Select the value from the dataset (source or target) that has fewest NULLs across all entities for the respective attribute.}   \\         	                
	 \hline     
\end{tabular}
%\vspace{.5em}
\newline
	* - requires metadata, A - All, S - String, C - Category (i.e., domain values have no order), T - Taxonomy (i.e., domain values have semi-order), N - Numeric. 

\label{tab:resolutionfunctions}
\end{table}

\begin{example} \label{exmp:strategyIV}
Applying strategy IV for synchronization on \autoref{ex:changeset} while resolving the conflicts using function 'Any' gives the following triples:

\begin{lstlisting}[breaklines=true]
dbr:Adrian_Sanders rdf:type      dbo:Politician;
		   rdf:type      dbo:Agent;
		   rdf:type      dbo:MemberOfParliment;
		   foaf:name     Adrian Sanders;
		   foaf:name     Sanders, Adrian;							  							  
		   dbp:birthYear 1959-01-01 (xsd:date);							  
		   owl:sameAs    http://wikidata.org/entity/Q479047;
		   owl:sameAs    http://yago-knowledge.org/resource/Adrian_Sanders.						
\end{lstlisting}
\end{example}

%
%\begin{lstlisting}[breaklines=true]
%dbr:FLIP_Burger_Boutique   a     dbo:Restaurant;
%         foaf:name         "FLIP burger boutique"@en;
%         georss:point      "33.7984 -84.4159";
%         dbo:abstract      "FLIP burger boutique (stylized as FLIP).."@en;
%         dbp:rating        4;
%         foaf:depiction    wmcf:Richard_Blais.JPG ;
%         dbp:headChef      "Richard Blais"@en.
%dbr:Jean_Georges  a        dbo:Restaurant;
%	 foaf:name         "JGV restaurant"@en;
%	 foaf:depiction    wmcf:FLIP.jpg;
%         georss:point      "40.76905277777778 -73.98143055555556";
%         dbo:abstract      "Jean Georges is a three-Michelin-stars restaurant at 1 Central Park ..."@en;
%         dbp:rating        5;
%         dbp:headChef      "Mark LaPico"@en.
%dbr:Het_Groot_Paradijs     a      dbo:Restaurant;
%         foaf:name         "Het Groot Paradijs";
%         dbo:abstract      "Het Groot Paradijs was a restaurant located in Middelburg,..";
%         georss:point      "51.50027222222222 3.6166805555555555";
%         dbp:rating        4.
%\end{lstlisting}
%\end{example}

\section{Approach}\label{sec:approach}
Our approach allows a user to choose a synchronization strategy (as presented in \autoref{def:strategies}). 
Below, we describe the status of the source and target datasets after applying each synchronization strategy (see \autoref{alg:datasetStatus}).
\begin{algorithm}[h] 
\KwData{$S_{t_i},T_{t_i}, \delta (T_{t_i-t_j}), \delta (S_{t_i-t_j} ), strategy$}
\KwResult{$S_{t_j},T_{t_j}$ }
\scriptsize
\Switch{strategy}{
 	\tcc{Synchronise with the source and ignore local changes}
 	\Case{Strategy I}{
 		$ T_{t_j} := T_{t_i} \cup \delta (S_{t_i-t_j} )$ \;
 		$ S_{t_j} := S_{t_j} $ \;
 	}
 	\tcc{Do not synchronise with the source and keep local changes}
 	\Case{Strategy II}{
 		$ T_{t_j} := T_{t_i}  \cup \delta (T_{t_i-t_j} )$ \; 
 		$ S_{t_j} := S_{t_i}  \cup \delta (S_{t_i-t_j} )$ \;  
 	}
 	\tcc{Synchronise with the source and target datasets and ignore conflicts}
 	\Case{Strategy III}{ 
 		$ S_{t_j}, T_{t_j} := CDR (\delta (S_{t_i-t_j} ), \delta (T_{t_i-t_j} ), T_{t_i}, false) $ \; 
 	}
 	\tcc{Synchronise with the source and target datasets and resolve the conflicts}
 	\Case{Strategy IV}{
 		$S_{t_j}, T_{t_j} := CDR (\delta (S_{t_i-t_j} ), \delta (T_{t_i-t_j} ), T_{t_i}, true) $ \; 
 	}
 }
\caption{Updating the source and target datasets by the chosen synchronization strategy.}
\label{alg:datasetStatus}
\end{algorithm}

Function $CDR$ is presented in \autoref{alg:CDR} which
(i) identifies conflicts for the case of strategy III and strategy IV, and then 
(ii) resolves conflicts only in case of strategy IV.
Our approach considers triple-based operations, explained below using seven cases, to identify conflicts. 
%Here, we specify triple-based operations and their effect on $ T_{t_j} $.
Consider three triples $x_1= ( s, \, p, \, o_1 )$, $x_2 = ( s, \, p, \, o_2 )$, and $x_3 = ( s, \, p, \, o_3 )$ which are in conflict with each other 
$x_1 \in \delta ( S_{t_i-t_j} )  \wedge x_2 \in \delta ( T_{t_i-t_j} ) \wedge x_3 \in \{\delta ( S_{t_i-t_j} )  \wedge \delta ( T_{t_i-t_j} )\}  \wedge o_1 \notequiv o_2  \notequiv o_3$. 
In the following we present seven cases of evolution causing conflicts. 
For the first three cases (I-III), the conflict resolution is straightforward. 
But for the cases IV-VII, we have to employ a conflict resolution policy to decide about  triples $x_1$ and $x_2$:

\begin{itemize}
\item \textbf{\emph{Case I:}} $x_1$ is added to $T_{t_j}$ if $x_1$ is added by the source dataset and $x_2$ is deleted from the target dataset: $ x_1 \in \delta (S_{t_i-t_j} )^+ \wedge x_2 \in \delta (T_{t_i-t_j}  )^- $.  
%\item \textbf{\emph{Case II:}} $x_2$ is added to $S_{t_j}$ if $x_1$ is deleted by the source dataset and $x_2$ is added to the target dataset: $ x_1 \in \delta (S_{t_i-t_j} )^- \wedge x_2 \in \delta (T_{t_i-t_j}  )^+ $.  
\item \textbf{\emph{Case II:}} $ x_1 $ is added to $ T_{t_j} $ if $x_1$ is modified by the source dataset and $x_2$ is deleted from the target dataset: $ x_1 \in \delta (S_{t_i-t_j} )^+  \wedge  x_2 \in \delta (S_{t_i-t_j} )^- \wedge x_2 \in \delta (T_{t_i-t_j}  )^- $.  
\item \textbf{\emph{Case III:}} $ x_2 $ is added to $ S_{t_j} $ if $x_1$ is deleted from the source dataset and $x_2$ is modified in the target dataset: $ x_1 \in \delta (S_{t_i-t_j} )^- \wedge x_2 \in \delta (T_{t_i-t_j}  )^+  \wedge x_1 \in \delta (T_{t_i-t_j}  )^- $.  
\item \textbf{\emph{Case IV:}}  if the triple $x_1$ is added to the source dataset and $x_2$ is added to the target dataset: $ x_1 \in \delta (S_{t_i-t_j} )^+ \vee x_2 \in \delta (T_{t_i-t_j}  )^+ $.
\item \textbf{\emph{Case V:}} if $x_3$ is modified by both source and target datasets: $ x_2 \in \delta (S_{t_i-t_j} )^+ \wedge x_3 \in \delta (S_{t_i-t_j} )^- \wedge x_1 \in \delta (T_{t_i-t_j}  )^+ \wedge x_3 \in \delta (T_{t_i-t_j}  )^- $.
\item \textbf{\emph{Case VI:}} if $x_1$ is modified by the target dataset: $ x_1 \in \delta (S_{t_i-t_j} )^+  \wedge x_2 \in \delta (T_{t_i-t_j}  )^+ \wedge x_1 \in \delta (T_{t_i-t_j}  )^- $. 
\item \textbf{\emph{Case VII:}} if $x_1$ is modified by the source dataset: $ x_2 \in \delta (S_{t_i-t_j} )^+  \wedge x_1 \in \delta (S_{t_i-t_j} )^- \wedge x_1 \in \delta (T_{t_i-t_j}  )^+ $. 
\end{itemize}

\begin{algorithm}[tb]
\label{alg:CDRALG} 
\KwData{$S_{t_i},T_{t_i}, \delta (T_{t_i-t_j}), \delta (S_{t_i-t_j} ), conflictresolution$}
\KwResult{$S_{t_j},T_{t_j}$ }
\scriptsize

$ T_{t_j} = \phi $ \;
$ S_{t_j} = \phi $ \;
$ temp   = \phi $ \;
 \tcc{$step_1$}
\For{all triples $ x_1 = (s_1,\, p_1,\, o_1) \in \delta (S_{t_i-t_j} )^+ $}{

 \tcc{finding triples which are in conflict with $x_1$}
$ X = \{ x_2 = (s_1,\, p_1,\, Node.ANY) \in \delta (S_{t_i-t_j} )^- \cup \delta (T_{t_i-t_j} )^+ \cup \delta (T_{t_i-t_j} )^- \cup T_{t_i} \}$ \;

  \If { $ X  == \phi $} {
	    $ temp =  temp \cup x_1 $ \;	 }	
   	    \Else 
   	   {  x = resolveConflict($ x_1, X $) \;
   	     $ temp =  temp \cup x $ \;}
}
%
%\For{all triples $ x_1 = (s_1,\, p_1,\, o_1) \in \delta (S_{t_i-t_j} )^- $}{
%\tcc{finding triples which are in conflict with $x_1$}
%$ X = \{ x_2 = (s_1,\, p_1,\, Node.ANY) \in \delta (S_{t_i-t_j} )^- \cup \delta (T_{t_i-t_j} )^+ \cup \delta (T_{t_i-t_j} )^- \cup T_{t_i} \}$ \;
%
%  \If { $ X  != \phi $} {
%	     x = resolveConflict($ x_1, X $) \;
%   	     $ temp =  temp \cup x $ \;}
%}	
 \tcc{$step_2$}
$ T_{t_i} := T_{t_i} \setminus \delta (T_{t_i-t_j} )^- \cup  \delta (S_{t_i-t_j} )^- $\;
$ S_{t_i} := S_{t_i}  \setminus \delta (T_{t_i-t_j} )^- \cup  \delta (S_{t_i-t_j} )^- $ \;
 \tcc{$step_3$}
$ temp :=  temp \cup \delta (S_{t_i-t_j} )^+ \cup \delta (T_{t_i-t_j} )^+$ \;
\tcc{Updating the target dataset}
$ T_{t_j} := T_{t_i} \cup temp $ \;
\tcc{Updating the source dataset}
$ S_{t_j} := S_{t_i} \cup temp $ \;
\caption{CDR algorithm: Conflict Detection and Resolution}
\vspace{-0.5em}
\label{alg:CDR}
\end{algorithm}

Algorithm~\ref{alg:CDR} shows the pseudocode of the procedure for updating the source and target datasets at the end of each timeframe.   	
The function \texttt{resolveConflict} identifies operations described in Case I-VII. 
In addition, for the cases IV-VII, it resolves conflicts based on the type of involved predicate.
As we discussed earlier, whether a conflict between two triple exists depends heavily on the type of property.
Consider two triples $(s,\, p,\, o_1)$ and $(s,\, p,\, o_2) $, if $p$ is \verb|rdfs:label|, we measure the similarity between $o_1$ and $o_2$ using the Levenshtein distance.
We pick both values of \texttt{rdfs:label} if their similarity is below a certain threshold otherwise we treat them as conflicting. 
The function \texttt{resolveConflict} identifies operations containing deleted in the source, deleted/added/modified in the target dataset. 
In case of deleted in the source dataset and added/modified by the target dataset, it returns a triple to be added in $T_{t_j}$ otherwise null. 

\begin{figure}[tb]
%\vspace{-2.15em}
\centering
\includegraphics[width=0.95\textwidth]{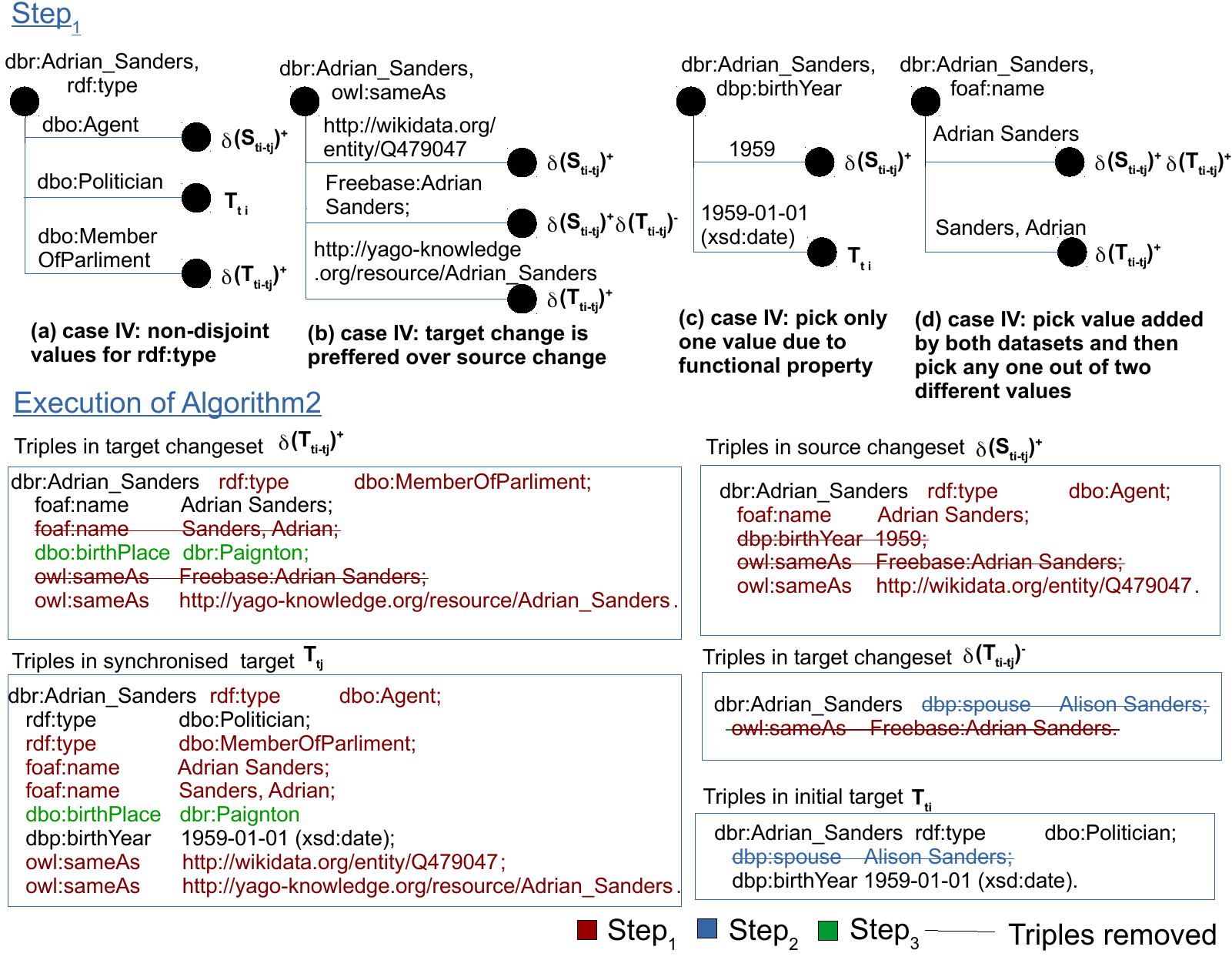}
\caption{Execution of \autoref{alg:CDR} to synchronize $ T_{t_i} $ with $S_{t_i}$}
\label{fig:DT2}
\vspace{-1.25em}
\end{figure}  

\autoref{fig:DT2} illustrates \autoref{alg:CDR} for updating the target dataset $T_{t_i} $. % to synchronize with source $S_{t_i}$ at some time point $ t_j $, using . 
We choose the synchronization strategy IV for the synchronization task.
In the first step, we use a tree structure to identify conflicts for the triples in $ \delta (S_{t_i-t_j} )^+ $. 
%Consider the tree structure (a) in $step_1$ for the triple $ (s_1,\, p_1,\, o_{11}) \in  \delta (S_{t_i-t_j} )^+ $. 
%We find different object values for $ (s_1,\, p_1) $ in $ \delta (S_{t_i-t_j} )^-, \delta (T_{t_i-t_j} )^+ $, $\delta (T_{t_i-t_j} )^-$,  and $ T_{t_i} $. 
%Then, we identify the triple based operation.
%For example, if we find the object value $ o_{11} $ in $ \delta (S_{t_i-t_j} )^+ , o_{12} $ in $ \delta (T_{t_i-t_j} )^+ $, and $ o_1 $ in $ \delta (S_{t_i-t_j} )^-, \delta (T_{t_i-t_j} )^- $, and $ T_{t_i} $ it means $  (s_1,\, p_1,\, o_1) $ is modified by both source and target. 
Consider the tree structure (a) in $step_1$ for the triple $(dbr:Adrian\_Sanders,\,$ $rdf:type,\, dbo:Agent) $. 
We find different object values for $ (dbr:Adrian\_Sanders,\, rdf:type) $ in $ \delta (S_{t_i-t_j} )^+, \delta (T_{t_i-t_j} )^+ $, and $ T_{t_i} $. 
Then, we identify the triple based operation.
For example, if we find the object value $ dbo:Agent $ in $ \delta (S_{t_i-t_j} )^+$ , $ dbo:MemberOfParliment $ in $ \delta (T_{t_i-t_j} )^+ $, and $ dbo:Politician$ in $ T_{t_i} $, it represents case IV of addition by both source and target. 
Thus, this case represents a potential conflicting triple. 
We check if the values in $T_{t_i}$, $\delta (S_{t_i-t_j} )^+$ and $ \delta (T_{t_i-t_j} )^+ $ are disjoint for predicate rdf:type.
As $ dbo:Politician $, $ dbo:Agent $, and $ dbo:MemberOfParliment $ are not disjoint, we pick all these values.
%We resolve it using the conflict resolution policy 'source preference' selected for predicate $ p_1 $ and add it in $ T_{t_j} $.   
%The triple-based operation for $step_1 (b)$ is source addition.  
%The $step_1 (c)$ for the triple $ (s_6,\, p_6,\, o_{60}) $ shows triple-based operations source addition and target modification.

Now, consider the tree structure (b) in $step_1$ for triple $(dbr:Adrian\_Sanders,\,$ $owl:sameAs,\, http://wikidata.org/entity/Q479047) $. 
It also represents case IV of addition by both source and target. 
The triple $(dbr:Adrian\_Sanders,\, owl:sameAs,\, Freebase:Adrian Sanders)$ is added by source but deleted by target. 
Considering the target as more customized dataset, we give preference to target change. 
The tree structure (c) in $step_1$ for the triple $(dbr:Adrian\_Sanders,\, dbp:birthYear,\, 1959) $. 
It is also handled in case IV. 
As dbp:birthYear is functional property, we select only one value among already existing value and the new value using resolution function 'Any'. 

Furthermore, the user has the opportunity to adopt the manual or automatic selection of resolution functions. 
The resolution function is oriented to the type of predicates. 
The list of supported resolution functions is shown in \autoref{tab:resolutionfunctions}.
For automatic selection of conflict resolution functions for predicates, we check attributes of predicates (e.g., type, cardinality). 
Based on the usage analysis of different functions in \cite{Bleiholder2006}, we prefer functions such as first, longest, and maximum for resolving conflicts.
For instance, we prefer function longest for strings to avoid loss of information. 
For numeric data types, we prefer function max to keep the up-to-date value.
For URIs, we pick the first value.

\section{Evaluation}\label{sec:evaluation}
%\vspace{-0.75em}
%To evaluate our proposed approach, we use the sample case study about politicians dataset and extract the data from DBpedia endpoint using the following SPARQL query.
In order to assess the discussed approaches for synchronization and conflict identification/resolution, we prepare a testbed based on a slice of DBpedia using the following SPARQL query.
\begin{lstlisting}
CONSTRUCT WHERE  {
    ?s    a                Politician ;
          foaf:name        ?name ;
          dbo:nationality  ?nationality ;
          dbo:abstract     ?abstract ;
          dbp:party        ?party ;
          dbp:office       ?office
    OPTIONAL { ?s  foaf:depiction  ?depiction }
}
\end{lstlisting}

The extracted dataset is used as the initial source and target dataset.
Then, we collect a series of changesets from DBpedia-live published from September 01, 2015 to October 31, 2015 using iRap~\cite{irap2015}.
We found a total of 304 changesets.
These changesets are leveraged to simulate updates of the source and target datasets.
We randomly select a total of 91 addition parts of changesets and altered values of their triples.
\autoref{tab:size} provides the number of triples of initial target, source and their associated changesets before synchronization.
Initially, we have \emph{200082} triples with \emph{163114} unique objects in $ T_{t_i}$ where $t_i  = September 01, 2015$. 

\begin{table}[ht]
\caption{ \scriptsize Number of triples in the source, target, and changesets for a given timeframe}
%\textbf{$T_{t_j}$}
\scriptsize
\centering
%\vspace{-1.5em}
 \begin{tabular}{|c|c|c|c|c|c|}
    \hline \textbf{$S_{t_i}$} & \textbf{$T_{t_i}$} & \textbf{$\delta (S_{t_i-t_j} )^+$} & \textbf{$\delta (S_{t_i-t_j} )^-$} & \textbf{$\delta (T_{t_i-t_j} )^+$} 
    &  \textbf{$\delta (T_{t_i-t_j} )^-$}  \\
        \hline 200082 & 200082 & 948 & 160 & 11725 & 81  \\     
    \hline 
\end{tabular}
%\vspace{-0.5em}
\label{tab:size}
\vspace{-2.15em}
\end{table}

%We execute the approach for sample data in five different scenarios.
Given a timeframe $t_i-t_j$ \footnote{09/01/2015-10/31/2015.}, the goal is to synchronize source and target datasets.
To do that, we define five different scenarios.
In four scenarios, we apply subsequently the strategy (I-IV) over all predicates of the changesets and measure the performance.
For the last scenario, we apply two strategies in a combined form on the changesets where we select strategy IV for predicate {\it dbp:office}, and strategy I for predicates {\it dbp:party}, {\it dbo:nationality}, 
{\it rdf:type}, {\it foaf:name}, {\it dbo:abstract}, and {\it foaf:depiction}.
%where we select strategy I for predicate dbp:party, strategy II for predicate dbo:nationality and dbp:office, and strategy IV for predicates rdf:type, foaf:name, dbo:abstract, and foaf:depiction.
For all predicates using strategy IV, we select the resolution function 'any'.
\autoref{tab:strategy} provides the number of triples produced as a result of synchronizing $S_{t_i}$ and $T_{t_i}$ in each scenario. 
%Changes $\delta (S_{t_i-t_j})^+$ and $\delta (S_{t_i-t_j})^-$ are sent back to source for synchronisation. 
The updated changesets are sent back to the source and target for synchronization purpose. 
The number of conflicting triples found in scenarios 3, 4, and 5 are shown in \autoref{tab:strategy}. 
\vspace{-2.5em}
\begin{table}[ht]
\caption{Results of synchronization}
\scriptsize
\centering
%\vspace{-1.5em}
 \begin{tabular}{|c|c|c|c|c|c|c|c|}
    \hline \textbf{Scenario} & \textbf{$\delta (S_{t_i-t_j})^+$} & \textbf{$\delta (S_{t_i-t_j})^-$} & \textbf{$\delta (T_{t_i-t_j})^+$}  
    & \textbf{$\delta (T_{t_i-t_j})^-$} & \textbf{Conflicting triples} & \textbf{RunTime (seconds)}\\
        \hline  1 & 0 & 0 & 948 & 160 & - & 0.0 \\    
        \hline  2 & 0 & 0 & 11725 & 81 & - & 0.0\\    %2 & 11725 & 81 & 11725 & 81 & - \\ 
        \hline  3 & 11682	& 81 &	12060 & 81 & 343 & 0.5\\     
        \hline  4 &  11800 & 195	& 12186 & 81 & 343 & 2.0\\       
        \hline  5 & 5227	& 131	& 6081 & 121 & 186 & 0.2\\     
        %\hline  5 & 10657	& 81	& 10732 & 89 & 8 \\                                 
    \hline            
\end{tabular}
%\vspace{-0.5em}
\label{tab:strategy}
\vspace{-2.5em}
\end{table}

The running time of the five different scenarios is also shown in \autoref{tab:strategy} (These times are recorded only for the execution of synchronization part and do not include data loading time). 
%Scenario 4 where conflicts for all predicates were found and resolved using strategy IV took more time as compared to all other scenarios.
Evaluation showed that strategy IV (performed in scenario IV) needs more time even from strategy III (performed in scenario III) where all conflicts were detected but not resolved. 

%\begin{table}[ht]
%\scriptsize
%\centering
%\vspace{-1.5em}
% \begin{tabular}{|c|c|c|c|c|c|c|}
%    \hline \textbf{Scenario} \\
%        \hline  1 & 0.0 \\    
%        \hline  2 & 0.0 \\    
%        \hline  3 & 0.5 \\     
%        \hline  4 & 2.0\\       
%        \hline  5 & 0.2\\                                  
%    \hline 
%             
%\end{tabular}
%\caption{Runtime per scenario}
%\vspace{-0.5em}
%\label{tab:time}
%\vspace{-2.05em}
%\end{table}
%\vspace{-1.05em}
Synchronization influences data quality specially in terms of data consistency.
To evaluate the usefulness of the synchronization approach, we use three data quality metrics i.e. (1) \emph{completeness}, (2) \emph{conciseness}, and (3) \emph{consistency} described as follows:
\begin{enumerate}
\item Completeness refers to the degree to which all required information is present in a dataset \cite{Zaveri2015}.
%Completeness defines how much data of a particular type exists in the dataset. 
We measure it for source and target changesets to identify which helps more in completeness. We measure it using
\vspace{-0.75em}
%\[ \frac{|\#\,of\,unique\,triples\,in\,synchronised\,dataset|}{|\#\,of\,unique\,triples\,in\,(initial\,dataset\,+\,changeset)|} \]
\[ \frac{Number\,of\,unique\,triples\,in\,synchronised\,dataset}{Number\,of\,unique\,triples\,in\,(initial\,dataset\,\cup\,changeset)} \]
\vspace{-0.75em}
\item Consistency states that the values should not be conflicting. We measure it using 
\vspace{-0.75em}
%\[ \frac{|\#\,of\,non-conflicting\,triples\,in\,synchronized\,dataset)|} {|\#\,of\,triples\,in\,(initial\,dataset\,+\,source\,and\,target\,changesets)|} \]
\[ \frac{Number\,of\,non\textsc{-}conflicting\,triples\,in\,synchronized\,dataset} {Number\,of\,triples\,in\,(initial\,dataset\,\cup\,source\,and\,target\,changesets)} \]
%- triples with functional object properties in (initial dataset + source changeset + target changeset)
\vspace{-0.75em}
\item Conciseness measures the degree to which the dataset does not contain redundant information using
\vspace{-0.75em}
\[ \frac{Number\,of\,unique\,triples\,in\,dataset}{Number\,of\,all\,triples\,in\,dataset} \]
%\vspace{-0.5em}
\end{enumerate}
Conciseness (before synchronization) is computed using initial target dataset and source and target changesets. 
We compute these metrics for all the assumed scenarios, the results are shown in \autoref{tab:complete}. 
For our sample case study, we found almost equal contribution of both source and target changesets in reducing the missing information.
However, we found minimum \emph{163191} number of unique objects using strategy II and maximum \emph{163591} number of unique objects using strategy IV.
Please note that strategy 1 and strategy II may not necessarily increase the number of unique triples as they do not consider about conflicts.  
It can be observed by analyzing the scenario 1 where the role of source changesets in completeness is 99\% which is less than the target contribution. 
Through evaluation, we found significant increase in conciseness for all strategies.

\begin{table}[h]
\caption{Synchronization effect on completeness, consistency, and conciseness}
\scriptsize
\centering
%\vspace{-1.5em}
 \begin{tabular}{|c|c|c|c|c|c|c|}
    \hline \textbf{Scenario} & \parbox{0.17\linewidth}{\textbf{Completeness (source)}} & \parbox{0.17\linewidth}{\textbf{Completeness (target)}} & \textbf{Consistency}  & \parbox{0.17\linewidth}{\textbf{Conciseness (before synchronization)}} & \parbox{0.17\linewidth}{\textbf{Conciseness (after synchronization)}} \\
        \hline  1 & 99\% & 100\% & - & 77\% & 81\% \\     
        \hline  2 & 99\% & 99\% & - & 77\% & 81\% \\   
        \hline  3 & 99\% & 100\% & 94\% & 77\% & 81\% \\     
        \hline  4 & 99\% & 100\% & 94\%& 77\% & 81\% \\       
        \hline  5 & 99\% & 100\% & - & 77\% & 81\% \\                               
    \hline             
\end{tabular}
%\vspace{-0.5em}
\label{tab:complete}
\end{table}
\vspace{-2.85em}

\section{Related Work}\label{sec:relatedwork}
Related work includes synchronization of semantic stores for concurrent updates by autonomous clients \cite{Aslan2011}, synchronization of source and target \cite{Tummarello2007}, replication of partial RDF graphs \cite{Schandl2010}, ontology change management \cite{Konstantinidis2007}, and conflict resolution for data integration \cite{Bleiholder2006, Motro2006, Michelfeit2014, Knap2012, Paton2012, Mendes2012, Schultz2011, Bilke2005, Bryl2014}. %Yan1999, Yager2004
We discuss related work here along the dimensions change management and conflict resolution.

\subsection{Change management}
Efficient synchronization of semantic stores is challenging due to the factors, scalability and number of autonomous participants using replica. 
\emph{C-Set}~\cite{Aslan2011} is a Commutative Replicated Data Type (CRDT) that %ensures consistency irrespective of the order of operations insert/delete at reception. 
allows concurrent operations to be commutative and thus, avoids other integration algorithms for consistency. 
%C-Set can be integrated within a semantic store and provides synchronization of autonomous semantic stores in a peer-to-peer network.
%\emph{RDFSync}~\cite{Tummarello2007} is a synchronization algorithm. 
%It converts an RDF graph into minimum self-contained graphs (MSGs) that contain minimal subsets of triples. 
%MSGs are represented by ordered list of unique identifier. 
%RDFSync computes the differences between the source and target ordered lists. 
%It offers three types of synchronization: 
%1) the target graph is equivalent to merge of source and target graphs, 
%2) the target removes unknown information to/from the source, and 
%3) the target graph is equivalent to source.\\
The approach, proposed in \cite{Schandl2010}, allows to replicate part of an RDF graph on clients. 
Clients can apply offline changes to this partial replica and write-back to original data source upon reconnection. 
%The partial replica also contains triples that describe which parts of the source RDF graph are not included in it. 
%For the purpose, the partial replica contains binary strings which are mapped to the triples in the source RDF graph. 
%This additional information helps to identify the changes on the client side upon reconnection.
%The coordination-free protocol~\cite{Ibanez2014} namely, Col-Graph, solves the writability, availability, and scalability issues of linked data (LD). 
%It allows LD consumers to replicate subset of data locally, perform querying, and modify replica to improve data quality. 
%LD consumer can make availability of replica to other participants through a SPARQL Endpoint. 
%The modifications can be sent back to the source data provider.
\autoref{tab:changemanagement} provides a comparative analysis of change management approaches used for synchronization.

A few surveyed approaches~\cite{Konstantinidis2007, Auer2006} are related to ontological change management.%, Papavassiliou2009
In \cite{Konstantinidis2007}, a framework is developed for ontology change management and tested for RDF ontologies. 
This framework allows to design ontology evolution algorithms. 
%It breaks the design of an ontology evolution algorithm into five steps: 
%1) selection of ontology representation model, 
%2) identification of change operations allowed on the ontology, 
%3) selection of consistency model, 
%4) resolution of inconsistencies, and 
%5) selection of most preferable actions determined in previous steps. 
In \cite{Auer2006}, an approach for the versioning and evolution of ontologies, based on RDF data model, is presented. 
%With support for versioning, it keeps track of different versions of an ontology while, with ontology evolution support, it identifies the changes leading to different versions. 
It considers atomic changes such as addition or deletion of statement and then aggregates them to compound changes to form a change hierarchy. 
This change hierarchy allows human reviewers to analyze at various levels of details. 
%In \cite{Papavassiliou2009}, a formal language of changes and a framework to define changes for RDF/S ontologies is developed. 
%The change language aggregates several low-level changes into high-level changes to form concise and intuitive changesets. 
%It associates all the triples of a low-level delta with only one high-level change which ensures that the change detection algorithm will handle changesets in a deterministic manner.
\vspace{-0.5em}
\begin{table}[b]
\caption{Synchronization approaches}
\centering
\scriptsize
 \begin{tabular}{|c|c|c|c|c|}
    \hline \textbf{Approach} & \textbf{Synchronization} & \textbf{Bi-directional} & \textbf{Participants} & \textbf{Conflict handling*}  \\
        \hline  \emph{C-Set} & \checkmark & \checkmark & n & x \\    
         \hline  \emph{RDFSync} & \checkmark & x & source, target & x \\
          \hline  \emph{Col-graph} & \checkmark & \checkmark & n & x \\
           \hline  [14] & \checkmark & back to source & n & x \\
            \hline  \emph{Co-evolution} & \checkmark & \checkmark & source, target & \checkmark \\
    \hline 
\end{tabular}
\newline
%\vspace{-0.5em}
	* - Triple level conflicts according to \autoref{def:conflict}
\label{tab:changemanagement}
\end{table}

\subsection{Conflict resolution}
%Data from multiple sources is integrated to increase the usefulness and quality of individual datasets. 
%Data integration is performed in multiple steps, one of which is data fusion where records representing same real world entity are combined to form a single and consistent representation~\cite{Bleiholder2006, Motro2006}.
%In \cite{Michelfeit2014}, a data fusion algorithm is proposed for the Linked Data integration framework \emph{ODCleanStore}~\cite{Knap2012}. 
%It resolves conflicts at schema, data, and identity level. 
%In contrast to our approach, it allows to resolve conflicts at query time like pay-as-you-go approach~\cite{Paton2012}. 
%Additionally, it computes quality scores of integrated data to help users to decide about the trustworthiness of results. 
%\emph{Sieve}~\cite{Mendes2012} is a data fusion module of \emph{LDIF}~\cite{Schultz2011}. 
%It uses quality scores to perform conflict resolution at the time of data loading into a data store. 
%It uses fusion functions PassItOn, KeepFirst, KeepLast, KeepAllValuesByQualityScore, Average, Maximum, Voting, and WeightedVoting. 
%For example, fusion function KeepFirst can select the value which belongs to a data source with highest reputation.

For relational databases, there is much work on inconsistency resolution \cite{Motro2006, Bilke2005, Bleiholder2006}. %, Yager2004, Yan1999
The \emph{Humboldt Merger}~\cite{Bilke2005}, extension to SQL with a FUSE BY statement, resolves conflicts at runtime. % by offering a variety of resolution functions including recency, voting, source preference, voting, or aggregation.
\emph{Fusionplex}~\cite{Motro2006} integrates data from heterogeneous data sources and resolves inconsistencies during data fusion. 
For fusion, it uses parameters such as user-defined data utility, threshold of acceptance, fusion functions, and metadata. %for example, maximum, average, or any, and source metadata including accuracy, recency, availability, or cost. 
%\emph{AURORA}~\cite{Yan1999} framework extends SQL to support conflict-tolerant queries. 
%Yager’s fusion framework~\cite{Yager2004} uses a voting-like mechanism to find the best value. 
\cite{Bleiholder2006} classifies conflict resolution strategies%and presents a catalog of their underlying resolution functions. % for use in an integrated information system~\cite{Bilke2005}. 
%It divides the conflict resolution strategies 
into three classes: ignorance, avoidance, and resolution. 
Conflict ignorance strategies are not aware of conflicts in the data. 
%For example, PassItOn and ConsiderAllPossibilities. 
Conflict avoidance strategies are aware of whether and how to handle inconsistent data. 
%These can be further divided into instance-based and metadata-based. 
%Source preferences can be based on reliability, cost, size, or some other quality criteria. 
Conflict resolution strategies may use metadata to resolve conflicts. 
These can be divided into deciding and mediating. 
A deciding strategy chooses value from already existing values whereas a mediating strategy may compute a new value. 

\emph{Sieve Fusion Policy Learner}~\cite{Bryl2014} uses a gold standard dataset to learn optimal fusion function for each property. 
The user specifies possible conflict resolution strategies from which the learning algorithm selects the one that gives maximum results within error threshold with respect to the gold standard. 

\begin{figure}[tb]
\centering
\vspace{-0.5em}
\includegraphics[width=0.7\textwidth]{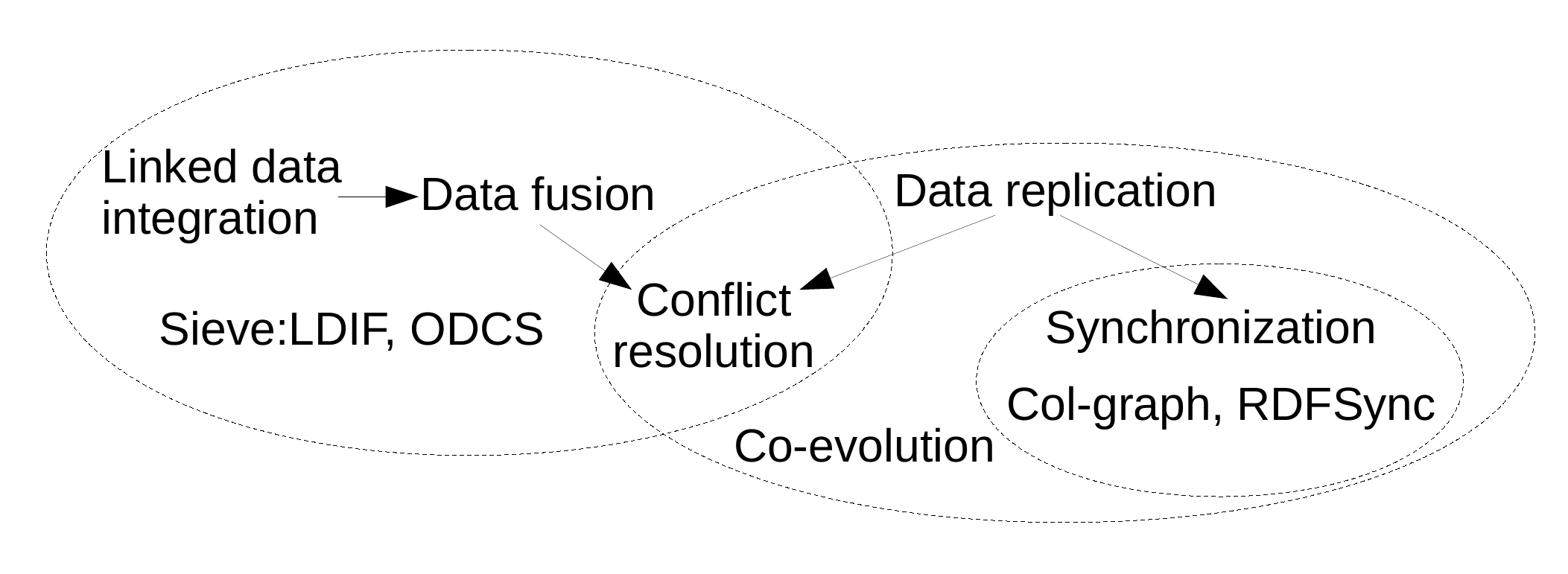}
\caption{How co-evolution fits with state-of-the-art}
\vspace{-0.5em}
\label{fig:related}
\end{figure}

Most relevant approaches to our proposed work are \emph{Sieve}~\cite{Mendes2012} - part of \emph{Linked Data integration framework (LDIF)}~\cite{Schultz2011}, data fusion algorithm~\cite{Michelfeit2014} for \emph{ODCleanStore}~\cite{Knap2012}, \emph{RDFSync}~\cite{Tummarello2007}, and \emph{Col-graph}~\cite{Ibanez2014}. 
Our approach differs from the previous ones in the scope of the problem (see \autoref{fig:related}).
RDFSync performs synchronization of two datasets by merging both graphs, deleting information which is not known by source, or making the target equal to source.
In contrast to RDFSync, our co-evolution approach allows merging of both graphs while ignoring or resolving conflicts and keeping only source or target changes.
Col-graph deals with consistent synchronization of replicas and does not tackle conflicts.

Sieve and ODCS are data fusion approaches and thus, are applicable where described data have different schemata. 
In contrast to both, co-evolution approach is applicable where described data have same schemata. 
Both approaches define conflicts as RDF triples sharing same subject/predicate with inconsistent values for objects. 
Sieve uses quality scores to resolve data while, ODCS produces quality scores of resolved data and keeps name of dataset from where the resolved value belongs.
We extend the conflict definition by further considering the predicate type, as discussed earlier (see \autoref{def:conflict}).

\section{Conclusion and Future Work}
\label{sec:conclusion}
In this paper we presented an approach to deal with co-evolution which refers to mutual propagation of the changes between a replica and its origin dataset.
Using the co-evolution process, we address synchronization and conflict resolution issues.
We demonstrated the approach using formal definitions of all the concepts required for realizing co-evolution of RDF datasets and implemented it using different strategies.
We evaluated the approach using data quality metrics completeness, conciseness, and consistency.
A thorough evaluation of the approach, using DBpedia changesets, indicates that our method can significantly improve the quality of dataset.
In the future, we will extend the concept of conflict resolution at schema level. 
For example, renaming a class invalidates all triples that belong to it in a dataset.
Further, we will evaluate the scalability and performance of our proposed approach using a benchmark dataset.
\newline
\newline
\textbf{Acknowledgements.} This work is supported in part by the European Union's Horizon 2020 programme for the projects BigDataEurope (GA 644564) and  WDAqua (GA 642795). 
Sidra Faisal is supported by a scholarship of German Academic Exchange Service (DAAD).

\bibliography{co-evolution}
\bibliographystyle{splncs03}

\end{document}